%% file: complexity.tex
\documentclass[aps,prl,twocolumn,groupedaddress,amssymb,amsfonts,showpacs]{revtex4}
\usepackage{times}

\usepackage{color}
\usepackage[dvips]{graphics,graphicx}

\input{ekqc_a}

\newcommand{\shortqph}[1]{}

\providecommand{\ignore}[1]{}

\newcommand{\mCo}[1]{\textcolor{blue}{}}

\newcommand{\hC}[1]{\textcolor{red}{}}

\def\proof{{\em Proof:  }}

\def\openone{\leavevmode\hbox{\small1\kern-3.8pt\normalsize1}}
\def\RR{{\rm I\kern-.2emR}}
\def\tr{{\rm tr}\; }

\def\fh{\mathfrak{h}}

\def\fsu{\mathfrak{su}}

\def\fso{\mathfrak{so}}
\def\proof{{\em Proof:  }}

\def\openone{\leavevmode\hbox{\small1\kern-3.8pt\normalsize1}}
\def\RR{{\rm I\kern-.2emR}}
\def\tr{{\rm tr}\; }

\def\ca{{\cal A}}

\def\cO{{\cal O}}

\def\hh{\hat{h}}
\def\hE{\hat{E}}
\def\hO{\hat{O}}
\def\hA{\hat{A}}

\def\hW{\hat{W}}

\def\hH{\hat{H}}

\def\hq{\hat{q}}
\def\he{\hat{e}}
\def\hX{\hat{X}}
\def\hY{\hat{Y}}
\def\hU{\hat{U}}
\def\hL{\hat{L}}

\providecommand{\ignore}[1]{}

\renewcommand{\ket}[1]{| #1 \rangle}
\renewcommand{\bra}[1]{\langle #1 |}

\newcommand{\bitem}{\begin{itemize}}
\newcommand{\eitem}{\end{itemize}}
\newcommand{\benum}{\begin{enumerate}}
\newcommand{\eenum}{\end{enumerate}}
\newcommand{\beq}{\begin{equation}}
\newcommand{\eeq}{\end{equation}}
\newcommand{\beqa}{\begin{eqnarray}}
\newcommand{\eeqa}{\end{eqnarray}}
\newtheorem{definition}{Definition}

\newtheorem{theorem}{Theorem}

\newtheorem{proposition}{Proposition}

\newtheorem{corollary}{Corollary}

\newcommand{\bproof}{\begin{proof}}
\newcommand{\eproof}{\end{proof}}
\newcommand{\bprop}{\begin{proposition}}

\newcommand{\bdef}{\begin{definition}}

\renewcommand{\cmplx}{{\mathbb C}}

\begin{document}


\title{Efficient solvability of Hamiltonians and limits on the power of some 
quantum computational models}



\author{Rolando Somma}
\affiliation{Los Alamos National Laboratory, MS D454, Los Alamos, NM 87545}
\email{somma@lanl.gov,barnum@lanl.gov,g_ortiz@lanl.gov}
\author{Howard Barnum}
\affiliation{Los Alamos National Laboratory, MS D454, Los Alamos, NM 87545}
\author{Emanuel Knill}
\affiliation{Mathematical and Computational Sciences Division, National
Institute of Standards and Technology, Boulder CO 80305}
\email{knill@boulder.nist.gov}
\author{Gerardo Ortiz}
\affiliation{Los Alamos National Laboratory, MS D454, Los Alamos, NM 87545}


\date{\today}
\begin{abstract}
We consider quantum computational models defined via a Lie-algebraic
theory. In these models, specified initial states are acted on by
Lie-algebraic quantum gates and the expectation values of Lie
algebra elements are measured at the end.  We show that these models can
be efficiently simulated on a classical computer in time polynomial in
the dimension of the algebra, regardless of the dimension of the
Hilbert space where the algebra acts. Similar results hold for the
computation of the expectation value of operators implemented by a
gate-sequence.  We introduce a Lie-algebraic notion of generalized
mean-field Hamiltonians and show that they are efficiently ({\it
exactly}) solvable by means of a Jacobi-like diagonalization method.
Our results generalize earlier ones on fermionic linear optics
computation and provide insight into the source of the power of the
conventional model of quantum computation.
\end{abstract}



\maketitle

Quantum models of computation are widely believed to be more powerful
than classical ones. Although this has been shown to be true in a few
cases, it is still important to determine when a quantum algorithm for
a given problem is more resource efficient than any classical one, or,
conversely, when a classical algorithm is just as efficient as any
quantum counterpart.  In general, one needs to know whether it is
worth investing in building a quantum computer (QC) and what is
required for success.  In this paper, we show close connections
between these issues and the efficient (or {\it exact}) solvability of
Hamiltonians. In particular, we show that a class of quantum models we
call generalized mean-field Hamiltonians (GMFHs)~\cite{sob04} is
efficiently solvable and furthermore does not provide a
stronger-than-classical model of computation: A quantum device
engineered to have dynamical gates generated by Hamiltonians from such
a set cannot directly simulate universal efficient quantum computation
and can be efficiently simulated by a classical computer (CC).

An algorithm is a sequence of elementary instructions that solves
instances of a problem.  It is said to be efficient if the resources
required to solve problem instances of size $N$ are
polynomial in $N$ ($\text{poly}(N)$) resources. Typically, the size of
a problem instance is the number of bits required to represent it, and
the relevant resources are time and space.  In the last
few years it has been shown that many pure-state quantum algorithms
can be efficiently simulated on a CC when the extent of entanglement
is limited (e.g., \cite{jl02,vid03}) or when the quantum gates available are far
from allowing us to build a set of universal
gates~\cite{gn97,val01,bsbn02}.  Here, we focus on a Lie algebraic
analysis to obtain other situations where quantum algorithms can be
efficiently simulated by CCs.  The so-called generalized coherent
states (GCSs)~\cite{per72} play a decisive role in our analysis.

The algorithms considered here make use of the Lie-algebraic model of
quantum computing (LQC). An LQC algorithm begins with the
specification of a semisimple, compact $M$-dimensional real Lie algebra $\hat{\fh}$ of
skew-Hermitian operators 
acting on a finite-dimensional Hilbert space $\cH$, with Lie bracket
$[\hX,\hY] := \hX \hY- \hY \hX$.
Without loss of generality, the action is irreducible. The
algorithm begins with a maximum-weight state $\ket{\sf hw}$ in $\cH$
and applies gates expressed as exponentials $e^{\hX}$ for certain
$\hX$ in $\hat{\fh}$.  The output of the algorithm is a noisy
expectation of an operator in $\fh$ or in $e^\fh$. LQC algorithms
cannot trivially be classically simulated because of the possibility
that the dimension of $\cH$ is exponential in the specification
complexity of $\hat{\fh}$ and $\ket{\sf hw}$. In order to precisely
define the model of LQC we require some results from the theory of Lie
algebras.  See~\cite{fuc92} for a textbook covering the basic theory
of Lie algebras.

Our intention is to restrict
observables and Hamiltonians to operators in $\sqrt{-1}\hat{\fh}$.
The dimension of $\cH$ may be exponential in $M$.  Since
we wish to implement computations with resources that are polynomial
in $M$, our knowledge of $\hat{\fh}$ cannot involve explicit matrix
representations of its operators.  We therefore assume that
$\hat{\fh}$ is specified as an abstract Lie algebra $\fh$ together
with a ``maximum weight'' $w$ characterizing its action on $\cH$.  For
computational purposes, we also use a small-dimensional faithful
representation of $\fh$. To be specific, we use the adjoint
representation, but for efficiency, one can choose the first
fundamental representation instead.  We use the following notational
conventions: Objects with a ``hat'' ($\hat{{\ }}$) belong to the
representation of $\fh$ on $\cH$. Objects with an ``overline''
($\bar{{\ }}$) belong to the chosen faithful representation.  Lie
algebraic objects with neither a hat nor an overline are associated
with the abstract Lie algebra (representation unspecified).
Implicit in these conventions are the representational isomorphisms
$\fh\rightarrow\hat{\fh}$ and $\fh\rightarrow\bar{\fh}$.

For the purpose of efficient representation, it is convenient to work
with the complexification $\cmplx\fh$ of $\fh$ and use a Cartan-Weyl
(CW) basis (see, for example, \cite{cor89}) for $\cmplx\fh$.  Thus, we
assume a decomposition $\cmplx\fh=\fh_D\oplus\fh^+\oplus\fh^-$, where
$\fh_D$ is a Cartan subalgebra (CSA), and $\fh^+$ and $\fh^-$ are
algebras of generalized raising and lowering operators,
respectively. $\fh_D$ is linearly spanned by named elements
$h_{1},\ldots h_{r}$, and $\fh^\pm$ by
$e^\pm_{\alpha_1},\ldots,e^\pm_{\alpha_l}$. The $\alpha_j$ are linear
functionals on $\fh_D$ called the positive roots of $\fh_D$.  The
abstract Lie algebra is specified by the identities $[h_k,h_l] = 0$,
$[h_k,e^\pm_{\alpha_j}] = \pm a_{jk}e^\pm_{\alpha_j}$,
$[e^+_{\alpha_j}, e^-_{\alpha_j} ] = \sum_k b_{kj} h_k$, and for
$j\not=k$, $[e^+_{\alpha_j}, e^\pm_{\alpha_{k}} ]= c_{jk}
e^+_{\alpha_j \pm \alpha_{k}}$. The bases of $\fh_D$ and $\fh^\pm$ may
be chosen so that the ``structure constants'' $a_{kj}$, $b_{kj}$ and
$c_{jk}$ are ratios of integers with $\text{poly}(M)$ digits.  The
structure constants do not uniquely specify the action of $\hat{\fh}$
on $\cH$. According to the representation theory of semisimple complex
Lie algebras, this action is uniquely specified by its ``maximum
weight'', which is a linear functional $w$ on $\fh_D$ given by its
values $w(h_k)$ on the distinguished basis of $\fh_D$.  The $w(h_k)$
are integral and are the eigenvalues of $\hh_k$ on the unique state
$\ket{\sf hw}$ annihilated by $\hat{\fh}^+$: $\hh_k\ket{{\sf hw}}=
w(h_k)\ket{{\sf hw}}$. The state $\ket{\sf hw}$ is called the ``maximum
weight state'' of the representation and its orbit under $e^{\hat\fh}$
generates the family of GCS of $\cH$ with respect to $\hat{\fh}$.  The
Hermitian inner product of $\cH$ and the Hermitian transpose of
operators on $\cH$ induce a corresponding Hermitian transpose
operation on $\cmplx\fh$.  We assume that the CW basis is chosen so
that the Hermitian transpose is given by $h_k^\dagger=h_k$ and
$(e^+_{\alpha_j})^\dagger=e^-_{\alpha_j}$.  We also assume that the
linear space on which $\bar{\fh}$ acts is endowed with a Hermitian
inner product for which the representation $\bar{\fh}$ is skew-Hermitian and
the Hermitian transpose matches the one defined for $\fh$.

The formal specification of an LQC algorithm requires the structure
constants of an abstract CW decomposition of $\fh$ and the weight
coefficients $w(h_k)$ determining $\ket{\sf hw}$.  The specification
complexities of $\fh$ and $\ket{\sf hw}$ are the number of bits
required to represent the numerators and denominators of the structure
constants and the $w(h_k)$.  Thus they are polynomial in the dimension
of $\fh$ and $\log\max(w(h_k))$.  
The {\it gates} of the algorithm may be
unitary exponentials $e^{t \hX}$, with $\hX$ a member of the CW basis.
The gate's resource requirement is the number of bits required to
represent $t$ plus $|t|$.  More generally, we can allow as gates any
$e^{\hH}$ with $\hH\in\fh$, where the resource requirement is given by
the specification complexity of $e^{\hH}$ (defined below).  There are
several alternatives for how the algorithm's output is obtained. We
consider two. In the first, the output is obtained by measuring the
expectation of an operator $\hA \in \cmplx\hat{\fh}$. In the second,
it is obtained from the absolute value of the expectation of an
operator $\hU\in e^{\cmplx\hat{\fh}}$.  The resource cost of making
the measurement is proportional to the sum of number of bits of precision and the
specification complexity of $\hA$ or $\hU$. The specification
complexity of $\hA$ is that of $A$ (the corresponding operator in the
abstract Lie algebra $\fh$) and is given by the number of bits used to
represent the coefficients of $A$ when expressed in the CW basis. If
$\hU$ is of the form $e^{\hH}$ with $\hH\in\cmplx\hat{\fh}$, its
specification complexity is that of $H$ plus $\max(|H_\alpha|)$ where
$H_\alpha$ ranges over the coefficients of $H$ expressed in the CW
basis.  Our assumption about the resource cost of measurement (i.e.,
the number of bits $b$ of precision) makes the LQC model just defined
very powerful but physically unreasonable. In particular, an LQC
algorithm gives exponentially better precision than an algorithm of
similar resource cost for the standard quantum computational model. The
standard quantum algorithm would need to be repeated exponentially many times in $b$
to return an expectation value with $b$ bits of precision.  In the
standard quantum computational model, the hypothetical ability to
determine expectation values with $b$ bits of precision using
resources polynomial in $b$ implies the ability to efficiently solve
problems in $\#P$, the class of problems associated with the ability
to count the number of solutions to NP-complete problems such as
satisfiability. This is a consequence of more general results
in~\cite{fenner:qc1999a}.

A natural question is when and how LQC can efficiently simulate, or be
simulated by, standard quantum or classical computation.  
The measurement models we introduced for LQC have the same form as
many typical problems in physics, which involve the evaluation of
correlation functions
\begin{equation}
\label{correl1}
\langle \hW \rangle = {\sf Tr} [\rho \hW],
\end{equation}
where $\rho=\sum_{s=1}^L p_s \ket{\phi_s} \bra{\phi_s}$ is the density
operator of the system ($p_s > 0; \sum_s p_s =1$), $\ket{\phi_s}$ are
pure states, and $\hW$ is a Hermitian or unitary operator acting on
$\cH$.  In general, the dimension $d$ of $\cH$ increases exponentially
in the problem size $N$, where the problem size is determined by
quantities such as the volume or number of particles of the system. An
algorithm to evaluate $\langle \hW \rangle$ with accuracy $\epsilon$
is efficient if the amount of resources required is bounded by
$\text{polylog}(d)+\text{poly}(1/\epsilon)$.

An efficient quantum algorithm to evaluate Eq. (\ref{correl1}) exists
if the state $\rho$ (or a good approximation to it) can be efficiently
prepared on a QC and if $\hW$ can be efficiently measured by using,
for example, the indirect techniques described in
Refs.~\cite{ogk01,sog02}. Unfortunately, known classical algorithms
for this purpose typically require resources polynomial in the
dimension $d$, which can be exponential in the problem size $N$. However,
if the problem can be specified Lie algebraically, this classical complexity can
be greatly reduced and exponential rather than polynomial
accuracy is efficiently achievable.

\begin{theorem}
\label{thm:1}
With $\rho$ as defined following Eq. (\ref{correl1}), 
if $\ket{\phi_s}=e^{\hA_s} \ket{\sf hw}$ are
GCS's of $\hat{\fh}$ ($\hA_s\in\hat{\fh}$) and $\hW\in \mathbb{C} \fh$,
then $\langle \hW\rangle$ can be classically computed to accuracy
$\epsilon$ in time polynomial in $log(1/\epsilon)$ and the sum of the
specification complexities of $\fh$, $\ket{\sf hw}$, $W$, $A_s$ and
$p_s$.
\end{theorem}

\proof We have $\langle \hW \rangle = \sum_{s=1}^L p_s \bra{{\sf hw}}
\hW_s \ket{{\sf hw}}$, where $\hW_s=e^{-\hA_s} \hW e^{\hA_s}$.  In
the CW basis,
\begin{equation}
\label{decomp1}
W_s = \sum\limits_{k=1}^r u^s_k \hh_k + \sum\limits_{j=1}^l v^{s+}_j
e^+_{\alpha_j} + v^{s-}_j e^-_{\alpha_j} , 
\label{GMFHdef}
\end{equation}
where $\ u^s_k,v^{s\pm}_j \in \mathbb{C}$.  To obtain these
coefficients, we can compute $\bar{W}_s$ in the adjoint
representation: $\bar{W}_s = e^{- \bar{A}_s} \bar{W} e^{\bar{A}_s}=
\sum\limits_{k=1}^r u^s_k \bar{h}_k + \sum\limits_{j=1}^l v^{s+}_j
\bar{e}^+_{\alpha_j} + v^{s-}_j \bar{e}^-_{\alpha_j}$.  To compute the
$u^s_k$ and $v^{s \pm}_j$ to accuracy $\delta$ requires computing the
matrix exponentials $e^{\pm\bar{A}_s}$, and matrix
multiplication followed by an expansion of the resulting matrix in
terms of the CW basis.  The matrix exponentials can be obtained to
accuracy $\delta'$ (in the $2$-norm) in time polynomial in
$\log(1/\delta')$ and the maximum of the entries of the $\bar{A}_s$ by
direct series expansion or other, more efficient
methods~\cite{ml03}. Matrix multiplication and basis expansion
increase the $2$-norm error by at most a constant factor, so that the
$u^s_k$ and $v^s_j$ can be efficiently obtained to the desired
accuracy.

Using the property that the $\he^{\pm}_{\alpha_j}$ either map
$\ket{\sf hw}$ to an orthogonal state or annihilate it, we
rewrite Eq.~(\ref{correl1}) as
\begin{equation} 
\label{correl3} \langle \hW \rangle = \sum_{s=1}^L p_s  
\sum\limits_{k=1}^r  u^s_k w(h_k)
\end{equation} 
and this sum can be evaluated efficiently with respect to the given
specification complexities. \qed

The following variant of Thm.~\ref{thm:1} holds for
$\hW= e^{\hH}$ with $\hH\in \mathbb{C}\fh$.
\begin{theorem}
\label{thm:1b}
If $\ket{\phi_s}=e^{\hA_s} \ket{\sf hw}$ ($\hA_s\in\hat{\fh}$) are GCS's of 
$\hat{\fh}$ and $\hW=
e^{\hH}$ with $\hH\in \cmplx\hat{\fh}$, then $|\langle \hW\rangle|^2$
can be classically computed to accuracy $\epsilon$ in time polynomial
in $log(1/\epsilon)$ and the sum of the specification complexities of
$\fh$, $\ket{\sf hw}$, $W$, $A_s$ and $p_s$.
\end{theorem}

\proof   
We can expand $|\langle \hW\rangle|^2$ as
\begin{eqnarray}
|\langle \hW\rangle|^2 =
  \sum_{s,s'}p_sp_{s'}
    \bra{\phi_s}\hW\ket{\phi_s}\bra{\phi_{s'}}\hW^\dagger\ket{\phi_{s'}}
        \nonumber\\
  = \sum_{s,s'}p_sp_{s'} \tr\hO_{s,s'},{\rm ~with~} \hO_{s,s'} := \nonumber \\     
\ket{{\sf hw}}\bra{{\sf hw}}e^{-\hA_s}e^{\hH}e^{\hA_s}\ket{{\sf hw}}
    \bra{{\sf hw}}e^{-\hA_{s'}}e^{\hH^\dagger}e^{\hA_{s'}}\ket{{\sf hw}}\bra{{\sf hw}} \;.
\nonumber
\end{eqnarray}
$\hO_{s,s'}$ is proportional to $\ket{{\sf hw}}\bra{{{\sf hw}}}$ and its trace
is the constant of proportionality. We can express $\ket{\sf
hw}\bra{{\sf hw}}$ as a limit of operators in $e^{\cmplx\fh}$.
Let $L=\sum_{k=1}^r w(h_k) h_k$ and define $\omega$ by $\hL \ket{{\sf
hw}}= \omega \ket{{{\sf hw}}}$. Then $\bra{\psi} \hL \ket{\psi} < \omega$ for
$\ket{\psi} \ne \ket{{{\sf hw}}}$, from which it follows that $\ket{{\sf
hw}} \bra{{{\sf hw}}} = \lim_{t \rightarrow \infty} e^{-t\omega} e^{t
\hL}$~\cite{bko03}.  Because the eigenvalues of $\hL$ are integral,
convergence is exponentially fast in $t$.  Let
\begin{equation}
E(t) = 
\sum_{s,s'}p_sp_{s'}
       e^{-3\omega t} e^{t L}e^{-A_s}e^{H}e^{A_s}e^{t L}
       e^{-A_{s'}}e^{H^\dagger}e^{A_{s'}}e^{t L}.
\end{equation}
$\hE(t)$ is positive definite Hermitian and converges to $\hO_{s,s'}$ as
$t\rightarrow\infty$.  For a given $t$, we can compute $\bar{E}(t)$ by
computing exponentials and multiplying matrices in the adjoint
representation. Observe that the maximum eigenvalue $\kappa(t)$ of
$\hE(t)$ converges exponentially fast to $|\langle\hW\rangle|^2$.  To
compute $\kappa(t)$ we first determine $\bar{Q}(t)$ such that
$\bar{E}(t) = e^{\bar{Q}(t)}$.  With the assumed Hermitian
inner product on the adjoint representation, $\bar{E}(t)$ is 
positive definite.  Thus, there is a unique Hermitian $\bar{Q}(t)$
satisfying $\bar{E}(t) = e^{\bar{Q}(t)}$, and $\bar{Q}(t)$ is
necessarily in $\sqrt{-1}\bar\fh$. The operator $\bar{Q}(t)$ can be
obtained via any conventional efficient diagonalization procedure for
non-negative definite matrices.  We can then use an efficient Jacobi-like
diagonalization procedure~\cite{wil93} to obtain unitary operators
$\bar U(t)\in e^{\bar{\fh}}$ and $\bar{q}(t) \in \bar{h}_D$ such
that $\bar U(t)\bar{q}(t)\bar U(t)^{\dagger}=\bar{Q}(t)$.  The maximum
eigenvalue of $\hE(t)$ is given by the exponential of the maximum
eigenvalue of $\hq(t)$.  At this point we require a number of results
from the representation theory of Lie algebras. For example,
see~\cite{fuc92}.  The element $q(t)$ induces an alternative order on the
roots, according to which a root $\alpha_j$ is positive if
$\alpha_j(q(t))$ is positive. (To remove degeneracies, it may be
necessary to slightly perturb $q(t)$.)  For this ordering, we
determine simple roots $\beta_k$ and corresponding members
$h'_k\in\fh_D$ such that $h'_k$ is isomorphic to
$h_k$ via a member of the Weyl group.
We can expand $\bar{q}(t)
= \sum_j q_j\bar{h}'_j$.  Uniqueness of maximal weights in
representations of Lie algebras implies that the maximum eigenvalue of
$\hat{q}(t)$ is given by $w'(q(t))=\sum_j w(h_k)q_j$.

We claim that the necessary steps can be implemented with polynomial
resources in the dimension of the Lie algebra and the number of digits
of precision of $\kappa(t)$. The matrix and root manipulations can be
implemented efficiently, but with respect to the precision of entries
of the matrix. It is necessary to realize that unless the weight $w$
is sufficiently small, $\bar{E}(t)$ converges to $0$ exponentially
fast in $t$. However, because the $|w(h_k)|$ are polynomial in $M$,
the rate of convergence to $0$ is bounded by $e^{-\text{poly}(M)}$. To
compute $\kappa(t)$ to a desired number $P$ of digits of precision, it
suffices to compute in the low dimensional matrix representation with
a precision of $\text{poly}(M)+\text{poly}(P)$ digits, which can still
be done with polynomial resources.  The relevant Weyl group
transformations can be done efficiently by use of one of the constructive
proofs of the transitivity of the Weyl group. See, for example, the
proof of Thm. 2.63 in~\cite{knapp:qc1996a}.  \qed

Important special cases motivating these results are {\it fermionic
linear optics} quantum computation (and equivalent {\it matchgate}
models introduced by Valiant), which is efficiently classically
simulatable \cite{val01, td02, kni01}, and models that also include
linear fermionic operators ($\fso(2N+1)$) for which an extension of
the canonical Bogoliubov mapping exists \cite{fuk77}. Natural bosonic
analogues of the fermionic results also exist. As a result it is
possible to efficiently simulate quantum computational models in which
coherent states are acted on by linear optical circuits, and measured
via homodyne detection \cite{bsbn02}, and of models with initial
multimode squeezed states and squeezing gates as well as linear ones
\cite{bs02}.  Like LQC with the second measurement strategy, these
involve the efficient simulation, in the dimension of a Lie algebra,
of a computational model in which coherent states of a Lie group with
gates generated by the algebra constitute the initial states and
computation. However, in the bosonic case the relevant algebra is not
semisimple, and the relevant irreps are infinite-dimensional.

We can now address the important question of the classical
simulatability of LQCs.

\begin{theorem}
\label{thm:2}
For both LQC measurement schemes, the result of an LQC algorithm $\ca$
can be obtained by use of classical computation in time polynomial in the
specification complexity of $\ca$.
\end{theorem}

\proof The action of the gates of the algorithm result in the state
$\ket{\phi} = \prod_{m=1}^t e^{\hA_m} \ket{{{\sf hw}}}$ where $\hA_m
\in \hat{\fh}$. Let
\begin{equation}
\label{correl4}
\langle \hW \rangle = \bra{{{{\sf hw}}}} \prod_{m=1}^t e^{-\hA_m} 
\hW \prod_{m=1}^t e^{\hA_m} \ket{{\sf
hw}}.
\end{equation}
The result of the algorithm is $\langle\hW\rangle$ if $\hW\in\fh$, or
$|\langle\hW\rangle|$ if $\hW\in e^{\cmplx\fh}$.  The result can be
computed by generalizing the algorithms given in the proofs of
Thms.~\ref{thm:1} and~\ref{thm:1b}. All that is required is to compute
the full product $\prod_{m=1}^t e^{\bar{A}_m}$ instead of the single
exponential required for Thms.~\ref{thm:1} and~\ref{thm:1b}. The
complexity of the method is polynomial with respect to the
specification complexity of $\ca$. \qed

The meaning of Thm.~\ref{thm:2} can be expressed in terms of
generalized entanglement~\cite{bko03,bko04,sob04}. A pure state is
generalized unentangled (GU) with respect to a preferred set $\cO$ of
observables if it is extremal among states considered as linear
functionals on $\cO$, otherwise it is generalized entangled. In a Lie
algebraic framework, a GU state is a GCS of a semisimple compact Lie
algebra. Thus, Thm.~\ref{thm:2} states that if a quantum computation
does not create generalized entanglement with respect to a
polynomial-dimensional semisimple compact Lie algebra, such a
computation can be efficiently simulated to exponential precision on a
CC.  We also remark that, because it cannot access all pure states
during the computation, but only the submanifold of generalized
unentangled ones, such a computation cannot directly simulate standard
quantum computation.

For applications to physics simulation, the following corollary shows
that higher-order correlation functions can also be computed
efficiently, provided the order is not too large.

\begin{corollary}
Let $\hW^1,\ldots,\hW^q$ be operators in $\cmplx\hat{\fh}$.
For fixed $q$, the expectation value of correlation functions of the
form $\langle \hW^1 \cdots \hW^q \rangle = \sum_{s=1}^L p_s \bra{{{\sf
hw}}} e^{-\hA_s} \hW^1 \cdots \hW^q e^{\hA_s}\ket{{{{\sf hw}}}}$,
can be computed on a CC in time polynomial in
$\log(1/\epsilon)$, and the sum of the specification complexities of
$\fh$, $\ket{{{\sf hw}}}$, $W^j$, $A_s$ and $p_s$.
\end{corollary}

The complexity of our algorithm for computing the correlation function
in the corollary is exponential in $q$.

\proof We outline an efficient algorithm for computing the desired
correlation function. First we expand each
$W^j_s=e^{-\hA_s}W^je^{\hA_s}$ in the CW basis as in the proof of
Thm.~\ref{thm:1}.  The desired correlation is given by
$\sum_s\bra{{\sf hw}}\prod_j W^j_s\ket{{\sf hw}}$.  We formally
multiply the CW basis expressions for the $W^j_s$ to obtain sums $P_s$
of formal products of members of CW basis representing the $\prod_j
W^j_s$.  Each product of CW basis members is standardized by using the
commutation rules so that each term is a product where all lowering
(raising) operators follow (precede) members of $\fh_D$.  This is
similar to the procedure of Wick's theorem. After this transformation,
terms that retain some lowering or raising operators contribute
nothing to the correlation functions. The remaining terms'
contribution is easily computed from $\bra{{\sf hw}}\hh_k\ket{{\sf
hw}} = w(h_k)$.  The contribution to the complexity of the procedure
of the formal multiplication and standardization procedure grows
exponentially in $q$. The number of terms that arise is bounded by
$\text{poly}(M)^q$, so that for fixed $q$, the complexity remains
polynomial in the given specification complexities. Further details
are available in~\cite{som05}.\qed

The algorithms given above can also be used to analyze certain
interacting physical models.  We use the term GMFH \cite{sob04} for
Hamiltonians belonging to ${\sqrt{-1}\fh}$ for $\fh$ in a sequence of
semisimple compact operator Lie algebras of dimension
$M\le\text{polylog}(d)$ acting on $d$-dimensional Hilbert spaces.  A
GMFH is necessarily specified in terms of a basis of $\fh$ that can be
efficiently transformed to a CW basis.  An example of a GMFH is given
by the $N$ spin-1/2 Ising model in a transverse magnetic field $\hH_I
= \sum\limits_{j=1}^N (g\sigma_x^j \sigma_x^{j+1} + \sigma_z^j)$,
where $\hH_I$ is an element of the Lie algebra $\fso(2N)$, with
dimension $M=2N^2 -N \equiv \text{polylog}(d)$, where $d=2^N$.
Interestingly, this model can be exactly solved and, as we will show,
this result can be extended to any GMFH.  We say that a Hamiltonian
acting on a $d$ dimensional Hilbert space can be efficiently ({\it
exactly}) solved when any one of its eigenvalues and a description of
the corresponding eigenstate can be obtained and represented to
precision $\epsilon$ in $\text{polylog}(d) + \text{poly}(1/\epsilon)$
computational operations on a CC.  In general, this definition makes sense
when we focus on Hamiltonians describing the interactions of $N$-body
systems, where $d$ increases exponentially with $N$.

\begin{theorem}
\label{theor:3}
GMFHs can be efficiently solved.
\end{theorem}

\proof 
Let $\hH_{MF}$ be a GMFH in ${\sqrt{-1}\hat{\fh}}$ given in
terms of a CW basis of $\fh$ as in Eq.~\ref{GMFHdef}.
We show that to solve $\hH_{MF}$ it suffices to diagonalize it
according to
\begin{equation}
\label{diagonal}
\hH_D=\hU \hH_{MF} \hU^\dagger = \sum_{k=1}^r \varepsilon_k \hh_k \ , 
\end{equation}
with $\varepsilon_k \in \mathbb{R}$ and $\hU\in e^{\hat{\fh}}$ unitary.  The
eigenvalues of $\hH_{MF}$ are shared with those of $\hH_D$. A
description of the corresponding eigenspaces consists of an eigenspace
of $\hH_D$ transformed by $\hU^\dagger$, where $\hU$ may be described
by a sequence of LQC gates.  According to the representation theory
of Lie algebras, the eigenspaces of $\hH_D$ consist of weight states
of $\hat{\fh}$, which can be obtained from the highest weight state by
applying lowering operators. They are characterized by linear
functionals $\lambda$ on $H_D$ of the form $\lambda(h_k) =
w(h_k)-\sum_{l} n_l \alpha_l(h_k)$, where the $n_l$ are non-negative
integers. Which choices of $n_l$ correspond to weight states is
readily determined from the representation theory of Lie algebras.
Once we have expanded $\hH_D=\sum_k \varepsilon_k\hh_k$, the eigenvalue
corresponding to $\lambda$ is readily computed as $\lambda(H_D) =
\sum_k \varepsilon_k\lambda(h_k)$.

To efficiently diagonalize $\hH_{MF}$ and obtain a specification of
$\hU$, we compute in the adjoint representation and apply a
generalization of the Jacobi method~\cite{wil93,khh04} to
$\bar{H}_{MF}$.  It yields an exponentially converging diagonalization
and an expression for $\hU$ in terms of a sequence of exponentials of
members of the $\fsu(2)$ subalgebras of $\cmplx\fh$ generated by the
pairs $\he^\pm_{\alpha_j}$. This suffices for our purposes.  \qed

{\em \bf Example.}  The fermionic Hamiltonians
$\hH_{MF}=\sum_{i,j=1}^N t_{ij} (c^\dagger_i c^{\;}_j -\delta_{ij}/2)
+ u_{ij} c^\dagger_i c^\dagger_j + h.c.$, where the operator
$c^\dagger_i$ ($c^{\;}_i$) creates (annihilates) a spinless fermion at
the $i$th site, belong to a representation of the Lie algebra
$\fso(2N)$ of dimension $M=2N^2-N \le \text{polylog}(d)$. A faithful
representation of $\fso(2N)$ is given by $c^\dagger_i
c_j-\delta_{ij}/2 \leftrightarrow T_{i,j}-T_{N+j,N+i}$, $c^\dagger_i
c^\dagger_j \leftrightarrow T_{i,N+j}-T_{j,N+i}$, and $c^{\;}_i
c^{\;}_j \leftrightarrow T_{N+i,j}-T_{N+j,i}$, where the $2N\times 2N$
matrices $T_{kk'}$ have +1 in the $k$th row and $k'$th column, and
zeros otherwise. Therefore, we write the matrix of $H_{MF}$ in this
representation and apply the Jacobi algorithm to diagonalize it. The
result is equivalent to the one given by the Bogoliubov
transformation~\cite{br86}, where the Hamiltonian maps as
$\hH_{MF}\rightarrow \hH_D = \sum_{k=1}^r \varepsilon_k (\gamma^\dagger_k
\gamma_k -1/2)$, where the operator $\gamma^\dagger_k$
($\gamma^{\;}_k$) creates (annihilates) a fermionic quasiparticle in
the $k$th mode.

Although LQC algorithms and GMFHs can be efficiently simulated or
solved on a CC, it may still be useful to implement the algorithms or
simulate GMFHs with QCs. In particular, there may be
problems where a key component is expressed in terms of LQC or GMFHs
but a more complex quantum computation is required to determine the
information of interest. One case of interest is where the LQC or GMFH
component requires preparing a GCS.  One way for such a GCS to
arise is as the ground state of a GMFH.  According to the next
theorem, such GCSs are efficiently preparable on a QC
that has efficient access to the LQC initial state and gates.

\begin{theorem}
Let a GCS $\ket{\phi}$ of $\hat{\fh}$ be specified as the ground state of a
Hamiltonian $\hH\in\sqrt{-1}\hat{\fh}$.  Then $\ket{\phi}$ can be prepared
by use of resources polynomial in the specification complexity of $H$ on
a QC with the ability to initialize $\ket{{\sf hw}}$ and efficiently
apply LQC gates.
\end{theorem}

\proof It suffices to determine a $\hU\in e^{\hat{\fh}}$ expressed as a
polynomial product of LQC gates such that $\hH=\hU\hH_D \hU^\dagger$ with
$\hH_D\in\hat{\fh}_D$ such that $D$ induces the root order associated with
the CW basis. (See the proof of Thm.~\ref{thm:1b} for how an
element of $\fh_D$ induces a root order.) The state $\ket{\phi}$ is
then obtained as $\hU\ket{{\sf hw}}$ and hence is efficiently preparable
using LQC operations. To determine $\hU$ we can first use the
generalization of the Jacobi method as discussed previously.  This
yields an element of $\fh_D$ that does not necessarily induce the
desired root order. To complete the determination of $\hU$ requires
using a sequence of Weyl reflections to obtain the desired root order.
The sequence may be obtained using the method mentioned at the end
of the proof of Thm~\ref{thm:1b}.
\qed

Our results provide analogues of the Gottesman-Knill
theorem~\cite{gn97} (cf. also \cite{nc00gktheoremref, ag04})
concerning the efficient simulatability of Clifford-group
computational models, and of results on the simulatability of certain
multimode coherent-state and squeezed-state computational models
\cite{bsbn02, bs02}.  One might hope for a treatment, perhaps based on
Lie groups and groups of Lie type, that will unify these results,
specifically those based on (1) finite dimensional semisimple Lie
algebras, (2) Bosonic linear optics with homodyne detection (tied to
an infinite-dimensional irreducible representation of a solvable Lie
algebra) and possibly squeezing (involving a nilpotent Lie algebra),
and (3) Clifford groups and semigroups. Our results cast additional
light on why quantum computers may be more powerful than classical
computers.  It is a crucial fact that the generators of its gate-set,
though their number can be chosen to grow polynomially, generate an
exponentially large Lie algebra acting on an exponentially large
Hilbert space.  If the growth of the dimension of the generated Lie
algebra is polynomial, a computation with this gate set using
compatible state preparations and measurements can be simulated with
polynomial efficiency on a classical computer by working in a
low-dimensional faithful representation of the Lie algebra. What other
algebraically constrained models of quantum computation are
efficiently classically simulatable?  Such structures may underlie the
efficient solvability of further classes of Hamiltonians of condensed
matter models, which go beyond the GMFHs, such as those solvable via a
Bethe-type {\em Ansatz}.

\acknowledgments We thank L. Gurvits for discussions and for pointing
out Ref. ~\cite{wil93} and the US DOE and NSA for support.
Contributions to this work by NIST, an agency of the US government,
are not subject to copyright laws.

\vspace*{-5mm}
\bibliography{complexity}

\end{document}

%% file: ekqc_a.tex
\usepackage{calc,ifthen}

\newlength{\elimdepthdim}
\newlength{\elimheightdim}
\newlength{\elimwidthdim}

\newlength{\strutdepthdim}
\newlength{\strutheightdim}
\newlength{\strutwidthdim}

\newcommand{\ket}[1]{\qvbar{#1}\qrangle}
\newcommand{\bra}[1]{\qlangle{#1}\qvbar}

\newcommand{\cH}{{\cal H}}

\newcommand{\cO}{{\cal O}}

\newcommand{\cmplx}{\mathbb{C}}

\newcommand{\proof}{\paragraph*{Proof.}}

\newcommand{\qed}{\hspace*{\fill}\rule{2.5mm}{2.5mm}%
\vspace*{8pt}\par}

\newcounter{herefignum}